\documentclass{article}
\usepackage{spconf,amsmath,graphicx}
\usepackage{bm}
\usepackage{amsfonts}

\title{SCA: Streaming Cross-attention Alignment for Echo Cancellation}
\name{Yang Liu, Yangyang Shi, Yun Li, Kaustubh Kalgaonkar, Sriram Srinivasan, Xin Lei}
\address{Meta, US \\
\{yangliuai, yyshi, yunli1, kaustubhk, sriramsri, leixin\}@meta.com}

\begin{document}
%
\maketitle
\begin{abstract}

End-to-End deep learning has shown promising results for speech enhancement tasks, such as noise suppression, dereverberation, and speech separation. However, most state-of-the-art methods for echo cancellation are either classical DSP-based or hybrid DSP-ML algorithms. Components such as the delay estimator and adaptive linear filter are based on traditional signal processing concepts, and deep learning algorithms typically only serve to replace the non-linear residual echo suppressor. This paper introduces an end-to-end echo cancellation network with a streaming cross-attention alignment (SCA). Our proposed method can handle unaligned inputs without requiring external alignment and generate high-quality speech without echoes. At the same time, the end-to-end algorithm simplifies the current echo cancellation pipeline for time-variant echo path cases. We test our proposed method on the ICASSP2022 and Interspeech2021 Microsoft deep echo cancellation challenge evaluation dataset, where our method outperforms some of the other hybrid and end-to-end methods.

\end{abstract}

\section{Introduction}

While the use of voice communication has seen rapid growth, cancelling acoustic echo without suppressing the near-end talker remains a major unsolved problem in providing high quality speech. Traditionally, digital signal processing (DSP) based linear echo cancellation has been applied based on estimating the acoustic echo path with an adaptive filter. This approach fails when the echo path is time-varying or non-linear which, results in either echo leaks or significant suppression of the near-end talker during double-talk.

In recent years, deep neural network (DNN) based acoustic echo cancellation (AEC) methods have achieved a significant improvement over the traditional signal processing based methods. Deep complex convolution recurrent network (DCCRN) designed for noise suppression \cite{hu2020dccrn} is modified for the AEC task with frequency-time LSTM (F-T-LSTM) network \cite{zhang2021ft} to better learn the relationship between frequency bands for effectively suppressing echo. The main drawback of inplace DC-CRN is the larger number of parameters. Recently, Indenbom et al. propose a self-attention alignment for AEC, which is capable of handling non-aligned microphone and far-end signals in linear and non-linear echo path scenarios \cite{indenbom2022deep}. In most cases, researchers assume that the echo path is linear and the time delay is limited to a known prior and effectively combine traditional signal processing with a neural network. Wang et al. use a deep feed-forward sequential memory network (DFSMN) as a post-filter after an adaptive filter based linear AEC \cite{wang2021weighted}. Peng et al. use a gated complex convolutional recurrent neural network (GCCRN) as a post-filter after adopted multiple filters \cite{peng2021acoustic}. However, the performance of the existing AEC algorithms, especially those with low complexity, may be greatly degraded in real-life practical applications. In these applications, software-related latency or hardware-related latency may lead to large time-variant delays. 

\begin{figure*}
    \centering
    \includegraphics[width=17cm]{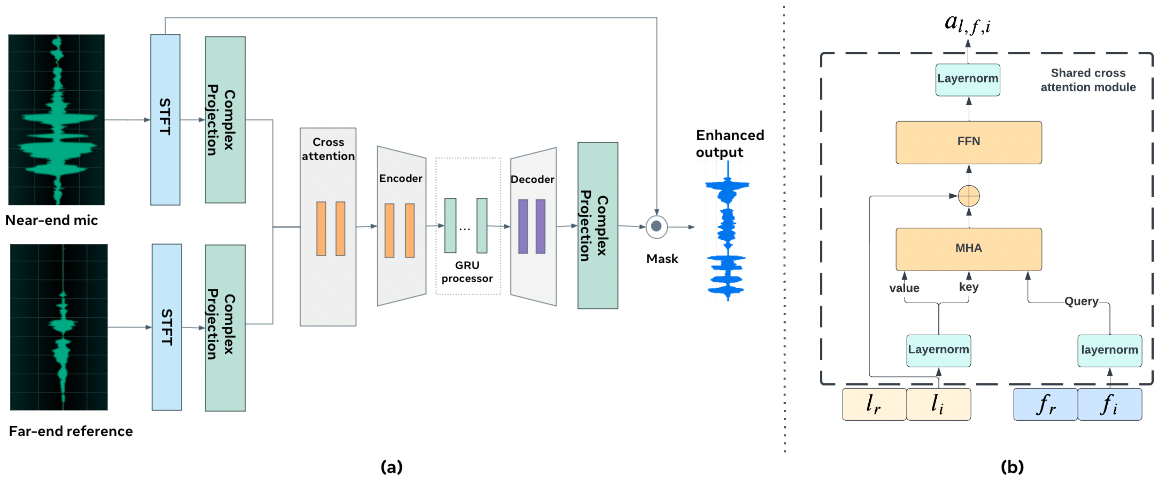}
    \caption{(a) streaming cross-attention alignment AEC network pipeline. (b) The calculation of the streaming cross-attention attention $a_{l,f,i}$. The cross attention from the imaginary $a_{l,f,i}$ and the real $a_{l,f,r}$ share the same cross-attention module. The cross attention from the far-end and near-end $a_{f,l,i}$ and $a_{f,l,r}$ uses another cross-attention module with same architecture.}
    \label{fig:my_label}
\end{figure*}

Inspired by the traditional signal processing alignment method such as cross-correlation \cite{ianniello1982time} and Emformer \cite{shi2021emformer}, we propose an end-to-end real-time streaming deep neural network without any extra alignment module. Compared to  prior works, our proposed streaming cross-attention alignment is used at the beginning of CRN network to improve the baseline model behaviour. The SCA-CRN has three contributions. First, cross-attention is applied to use near-end microphone signal to align the far-end signal. Compared to self-attention AEC \cite{indenbom2022deep}, this work adds multihead cross attention together with other components in transformer~\cite{Vaswani2017} like layer norm, feedforward neural (FFN) layer and projection operations. Multihead self-attention exploits the long range cross dependency between near-end signal and far-end signal. The feedforward neural network and projection operations are essential to generate a good representation from the multihead cross-attention output. Second, real and imaginary information are considered as two independent tensors to avoid complex computation, but the real and imaginary tensors share the weight of attention layers. Compared to complex-value attention used in \cite{indenbom2022deep}, SCA is straightforward to implement in real-time applications. Third, SCA is implemented with a streaming mask to limit the cross-attention to access very limited look-ahead context in training which supports the models for low-latency streaming applications.

\section{Proposed method } 
\subsection{Problem formulation}
For a generic AEC system, we define the microphone signal as $d(n)$ which consists of near-end speech $s(n)$, acoustic echo $z(n)$ and background noise $v(n)$:
\begin{equation}
d(n) = s(n) + z(n) + v(n), 
\end{equation}
where $n$ refers to the time sample index. $z(n)$ is a delayed version of the far-end reference signal $x(n)$ via echo path with potential nonlinear distortions caused by loudspeakers. This delay is related to the echo propagation path between the microphones and loudspeakers, hardware and software. This means the delay is time varied, unknown and difficult to estimate. The AEC task aims to separate $s(n)$ apart from $d(n)$, on the premise that ``unaligned" $x(n)$ is known. The error signal $e(n)$ and linear echo $y(n)$ are generated using $x(n)$ and $d(n)$ by using standard adaptive filtering techniques. 

\subsection{Network architecture}
The network structure is modified based on DC-CRN \cite{hu2020dccrn}. The SCA-CRN is composed by the SCA module, encoder, recurrent attention module and decoder, as shown in Figure 1(a). All audio signals are sampled at 48kHz. 

The network consists of two branches to be input into the SCA module. They are complex projection of the near-end microphone recording and the far-end reference recording. For running on edge devices, the models do not use complex tensors. The real and imaginary components are considered as two independent real tensors (see Section 3.2 for more details). The output of SCA is an embedding tensor consisting of far-end reference information and near-end microphone information and the input of the encoder. Each convolution block of the encoder is built by a gated convolutional layer \cite{dauphin2017language} which includes a gated linear unit (GLU) activation function and batch-norm layer. The number of kernels are 8, 16, 16 and all convolution kernels have a size of $2 \times 2$ with stride of $1 \times 2$ and we make the convolution layer causal by padding the incoming signal. At the decoder, each transpose convolution block is composed by a gated transposed convolution layer with GLU activation function followed by a batch norm layer. The number of kernels at the encoder are 16, 8, 2 and all convolution kernels have size $2 \times 2$. At the end the decoder, complex projection layer and gate mask layer
are applied. The estimated mask is applied to the near-end microphone signal. Each encoder and corresponding decoder layers are connected with a skip connection.

\subsection{Streaming cross-attention}
Assume that the complex projection outputs from near-end mic recording $d(n)$ and far-end reference input $x(n)$ are $[l_{r}, l_{i}]$ and $[f_{r}, f_{i}]$, respectively. The subscript $r$ denotes the real part and $i$ the imaginary part. Both $[l_{r}, l_{i}]$ and $[f_{r}, f_{i}]$ are stored as $\mathbb{R}^{b,t,c,d}$ where $b$ is the batch size, $t$ is the sequence length, $c$ is the 2 channels which are real channel and imaginary channel, and $d$ is the output dimension. 


The cross-attention uses a shared multiheads self-attention \cite{Vaswani2017} to explore the long term dependencies for both real and imaginary part. Given a pair input $l_{i}$ and $f_{i}$,  Fig~1(b) illustrates the way to get the cross attention output $a_{l, f, i}$. The detailed formulations are as follows:
\begin{align}
\vspace{-3pt}
&Q = \mathrm{LayerNorm}(f_{i}), \\
&K = \mathrm{LayerNorm}(l_{i}), \\
&V = \mathrm{LayerNorm}(l_{i}). 
\vspace{-5pt}
\end{align}
Based on the $Q, K, V$, the multihead attention\cite{Vaswani2017}  is applied
\begin{align}
\vspace{-3pt}
&\mathrm{Attention}(q, k, v) = \mathrm{Softmax}(\frac{\mathrm{Mask}(qk^{T})}{\sqrt{d_h}})v, \\
&head_{n} = \mathrm{Attention}(QW_{n}^Q, KW_{n}^K, VW_{n}^V), \\
&\mathrm{MHA}(Q, K, V) = \mathrm{Concate}(head_{1},..., head_{h})W^O,
\vspace{-5pt}
\end{align}
where $d_h$ is the dimension for each attention head. Assume the number of heads is $N$, then $d_h = \frac{d}{N}$. $W^O\in\mathbb{R}^{d, d}$ is the projection matrix for the attention output projection. $W_{n}^Q\in\mathbb{R}^{d, d_h}$, $W_{n}^K\in\mathbb{R}^{d, d_h}$, $W_{n}^V\in\mathbb{R}^{d, d_h}$ are the projection matrix for $Q$, $K$ and $V$ in each attention head, respectively. Similar to \cite{povey2018time,shi2021emformer}, the attention mask is used to limit the look-ahead context access for streaming, $\mathrm{Mask}(qk^{T})$ mask the product from query and key to be negative infinity, which essentially makes the attention weight to be zero after normalization from softmax.

The cross attention output $a_{l, f, i}$ is the result by feeding the multihead attention output through a residual connection, a feed forward network ($\mathrm{FFN}$) and a layer norm operation as follows:
\begin{align}
\vspace{-3pt}
a_{l, f, i} = \mathrm{LayerNorm}(\mathrm{FFN}(l_{i} + \mathrm{MHA}(Q, K, V)))).
\vspace{-5pt}
\end{align}

Similarly, we can get $a_{f, l, i}$ the cross attention between $f_i$ and $l_i$, $a_{f, l, r}$ the cross attention between $f_r$ and $l_r$, and $a_{l, f, r}$ the cross attention between $l_r$ and $f_r$. The output of the cross attention $[ca_r, ca_i]$ also store in shape of $\mathbb{R}^{b,t,c,d}$ containing two channels from real $ca_r$ and imaginary $ca_i$. Both $ca_r$ and $ca_i$ are the concatenation of the cross attention output from the real part and imaginary part.
\begin{align}
\vspace{-3pt}
ca_r = \mathrm{Concate}(a_{l, f, r}, a_{f, l, r}) \\
ca_i = \mathrm{Concate}(a_{l, f, i}, a_{f, l, i}).
\vspace{-5pt}
\end{align}


\subsection{Loss function}

We train the network with Mean squared error (MSE) loss on time domain and weighed MSE spectral loss on the magnitude spectrum. In the training stage, the enhanced near-end speech signal $\hat{s}(n)$ and target signal $s(n)$ are fed into the MSE loss function. Further, their complex spectrum of these signal processed by STFT and produced with weighting factor are fed into the weighted MSE spectral loss. Formally, the loss function is given by
\begin{equation}
\mathcal{L} = \alpha \sum_n |\hat{s}(n) -  s(n)| + \beta \sum_{n,k} w_k |S (\hat{s}(n)) -S ({s}(n)) |
\end{equation}
where the weighting factors $\alpha$, $\beta$ and  $w_k$ are heuristically determined to account for distortions in both the low and high frequency regions of the spectrum. The time and frequency indices as n and k for brevity and $S(.)$ is STFT function.

\begin{table*}[]\small
\centering
\begin{tabular}{c|c|c|c|c|c|c|c|c}
\hline
id    &    Model  & Aligned Input  &  Attention & RNN          & Size      & ERLE of FEST & PESQ of NEST & PESQ of DT \\ \hline
1& CRN          & No             & No & BLSM     &  7.8M           & 24.32      &   4.29        & 1.82      \\ \hline
2 & CRN   & Global GCC     & No& BLSM     & 7.8M                 & 33.27    &    4.55        &  2.46      \\ \hline
3 &NCA-CRN        & No             & NCA & BLSM     & 7.8M     &  40.17     &   4.54       &  3.03    \\ \hline
4 &NCA-CRN        & Global GCC     &  NCA & BLSM    & 7.8M     &  40.20      &  4.55        & 3.02    \\ \hline
5 & CRN        & Streaming GCC  & No& LCBLSTM                 & 7.8M     & 23.68      &   4.36       & 2.01       \\ \hline
6 & SCA-CRN        & No            &  SCA  & LCBLSTM      & 7.8M     &  32.17     &   4.50       & 2.60       \\ \hline
\end{tabular}
\caption{Performance comparison over candidate models. NCA: Non-streaming cross-attention alignment. SCA: Streaming cross-attention alignment. We measure WB-PESQ both DT and NEST scenarios and ERLE for FEST scenario in the augmented evaluation dataset.}
\end{table*}

\begin{table}[]\small
\centering
\begin{tabular}{l|l|l|l}
\hline
&  FEST &  DT & All  \\ \hline
Align-CRUSE &     4.46        &      4.56     & N/A   \\ \hline
GT-CrossNet &   N/A       &  N/A      &    4.29     \\ \hline
NCA-CRN  &     4.55    &   4.69     &  4.29     \\ \hline
SCA-CRN &     4.50      &   4.67       &  4.27     \\ \hline

\end{tabular}
\caption{AECMOS comparison against Align-CRUSE and GT-CrossNet baseline approaches. }
\end{table}

\section{Experimentation Results}

\subsection{Dataset and augmentation}
 We choose both the synthetic data from AEC-challenge \cite{cutler2022AEC} and our private augmented data to train the models. We balance the speakers' genders at both far-end and near-end sides and form total 720 original conversations with each 10s duration. The following typical use cases are considered to augment each conversation.

\textbf{Reverberation time (RT60)} the image method \cite{allen1979image} is used to produce both steady and time-variant room impulse responses (RIR) for creating echo paths in typical laptop settings. The RT60 is chosen to have probabilities of 0.6, 0.3, 0.08 and 0.02 over 50 $\sim$ 300 ms, 300 $\sim$ 600 ms, 600 $\sim$ 1 s and 1 $\sim$ 1.5 s. \textbf{Delay} between the playback and its received echo is introduced with probabilities of 0.05, 0.6, 0.4, 0.05 over -20 $\sim$ 0 ms, 0 $\sim$ 200 ms, 200 $\sim$ 400 ms and 400 $\sim$ 600 ms. \textbf{Signal-to-noise ratio (SNR)} is simulated by using typical noises from DNS-challenge \cite{dubey2022icassp} with probabilities of 0.1, 0.1, 0.3 and 0.5 over 0 $\sim$ 10 dB, 10 $\sim$ 20 dB, 20 $\sim$ 30 dB and 30 $\sim$ 40 dB; \textbf{Signal-to-echo ratio (SER) }is simulated with probability of 0.1, 0.5, 0.3 and 0.1 over -10$\sim$ 0 dB, 0 $\sim$ 10 dB, 10 $\sim$ 30 dB, 30 $\sim$ 40 dB;\textbf{Non-linearity} is simply modelled by either a arc-tangent to imitate gain saturation or a polynomial function as illustrated by \cite{zhang2022lcsm}; \textbf{Echo path changes} is introduced by randomly cutting or adding speech and silence segments of 10 to 200 ms to either near-end or far-end signals with a probability of 0 $\sim$ 10\%.

Each augmented conversation further converts to far-end single talk (FEST), near-end single talk (NEST) and double talk (DT) scenarios. The augmentation results in a total of 720K augmented conversations of roughly 2k hours. We use 70\% of this augmented data combined with the entire synthetic one from AEC-challenge for training and the rest 30\% of the augmented data for evaluation. 

\subsection{Ablation study}
Table 1 illustrates the performance of candidate models using non-streaming (model 1$\sim$4) and streaming (model 5-6) manners over the augmented evaluation data. The input of CRN-2, NCA-CRN-4 and CRN-5 has been aligned by generalised cross-correlation (GCC) \cite{benesty2004time}. Non-streaming cross-attention alignment (NCA) removes the mask $\mathrm{Mask}(qk^{T})$ of SCA to support non-streaming model. Compared with CRN-1, CRN-2 shows that the aligned inputs dramatically improve the AEC performance. With the non-streaming cross-attention mechanism, NCA-CRN-3 achieves significant improvement with the echo return loss enhancement (ERLE) and PESQ \cite{recommendation2001perceptual} increased by 20.7\% and 23.2\% respectively, as compared to CRN-2. NCA-CRN-4 shows that the performance remains almost the same with additional offline alignment introduced to NCA-CRN-3. Those evidences verify that NCA-CRN-4 is capable of replacing offline alignment as desired. CRN-5 and SCA-CRN are modified versions to support streaming processing without looking ahead. With the same model size, SCA-CRN improves ERLE and PESQ by 36\% and 29\% respectively. 

To illustrate that how cross-attention alignment handles data with different delays, we select a subset from the evaluation data where only delay variation and FEST scenario are considered. We choose CRN as baseline to compare with both NCA-CRN and SCA-CRN over ERLE performance with respect to the delays. Figure 2 shows that NCA-CRN suppresses the most echo with 5$\sim$20dB ERLE improvement from the baseline over the delay up to 400ms. SCA-CRN suppresses additional 5$\sim$7dB from the baseline within 200ms delay, which covers most of the normal use cases in voice communication. We exclude the non-causal delays in the comparison since SCA-CRN processes only previous data.       
\begin{figure}
    \centering
    \includegraphics[width=7cm]{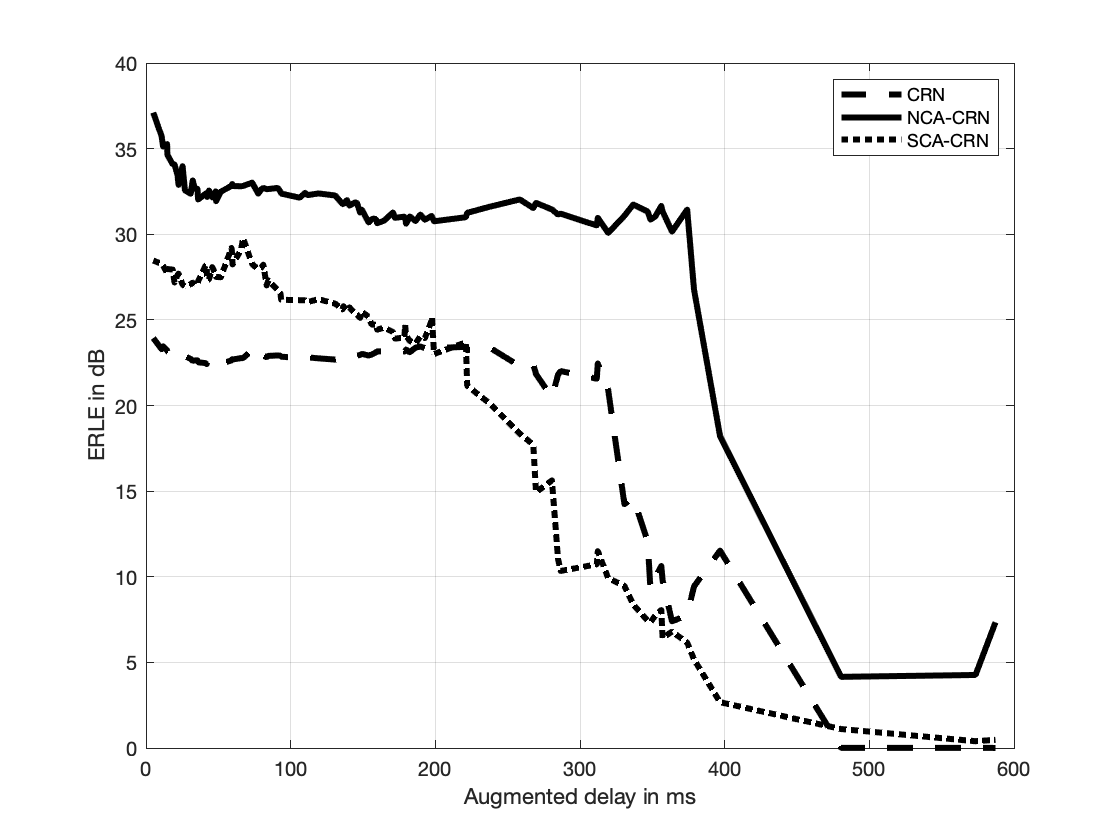}
    \caption{Comparison of unaligned modeling performance for FEST scenario vs augmented delays}
    \label{fig:my_label}
\end{figure}

\subsection{Comparison with the state-of-the-art methods}  
We use AECMOS, a non-intrusive model-based metric provided by AEC challenge, to compare our approaches with the chosen baselines Align-CRUSE \cite{indenbom2022deep} and GT-CrossNet \cite{zhao2022deep}. The blind test data from AEC challenge ICASSP2022 \cite{cutler2022AEC} is used to compare against GT-CrossNet and the counterpart from Interspeech2021 \cite{cutler2021interspeech} is used to compare against Align-CRUSE. Both data sets are real-world and the realistic distribution of delays has been demonstrated in \cite{indenbom2022deep}. Besides FEST and DT scenarios, in ``All" scenario, we take the average of all 4 types - FEST echo MOS, NEST other MOS, DT other and echo MOS to indicate the overall performance. Table 2 shows both NCA-CRN and SCA-CRN outperform Align-CRUSE in FEST and DT scenarios and are on par with GT-CrossNet in "All" scenario. From complexity perspective, SCA-CRN is less than half the model size of GT-CrossNet (17.4M) for data of 48k sampling frequency.

\section{Conclusion}

We proposed a novel streaming cross-attention and apply this attention on the convolution recurrent echo cancellation network. With multi-head cross-attention, layer norm, FFN layer and projection operations, the proposed SCA-CRN is able to handle unaligned input signals as well as other challenging echo scenarios, such as time-variant echo path, without additional costly alignment processing. The non-streaming version of SCA-CRN, named as NCA-CRN is proposed for non-streaming echo cancellation task. SCA-CRN and NCA-CRN both achieve encouraging improvement over real public data as compared with baseline approaches, especially in double talk scenarios. In future work, SCA can also be used to address multi-microphone alignment for speech enhancement.

\bibliographystyle{IEEEbib}
\bibliography{strings,refs}

\end{document}